\documentstyle[aps]{revtex}
\begin{document}
\centerline{\LARGE\bf What quantum mechanics describes is
discontinuous motion} \vskip 1cm \centerline{Gao Shan}
\centerline{Institute of Quantum Mechanics} \centerline{11-10,
NO.10 Building, YueTan XiJie DongLi, XiCheng District}
\centerline{Beijing 100045, P.R.China} \centerline{E-mail:
gaoshan.iqm@263.net}

\vskip 1cm
\begin{abstract}
We show that the natural motion of particles in continuous
space-time (CSTM) is not classical continuous motion (CCM), but
one kind of essentially discontinuous motion, the wave function in
quantum mechanics is the very mathematical complex describing this
kind of motion, and Schr\"{o}dinger equation is just its simplest
nonrelativistic motion equation, we call such motion quantum
discontinuous motion or quantum motion; furthermore, we show that,
when considering gravity the space-time will be essentially
discrete, and the motion in discrete space-time (DSTM) will
naturally result in the collapse process of the wave function,
this finally brings about the appearance of classical continuous
motion (CCM) in macroscopic world.
\end{abstract}

\vskip 1cm

\section{Introduction}
The analysis about motion has never ceased since the old Greece
times, but from Zeno Paradox to Einstein's
relativity\cite{Einstein}, only CCM is discussed, and its
uniqueness is taken for granted ever since, but as to whether or
not CCM is the only possible and objective motion, whether CCM is
the real motion or apparent motion, no one has given a definite
answer up to now, in fact, people have been indulging in the study
of the motion law, but omitted the study of the motion itself.

On the other hand, we have entered into the microscopic world for
nearly one century, but our understanding about it is still in
confusion, the orthodox view\cite{Bohr} renunciates CCM in
microscopic world, but permits no existence of objective motion
mode for the microscopic particles, while the
opponents\cite{Bell,Bohm,Everett} still recourse to CCM to lessen
the pain of losing realism, no other objective motion modes have
been presented for the microscopic particles till now, thus the
above problem is more urgent than ever.

In this paper, we will mainly address the above problem, after
given a deep logical and physical analysis about motion, we
demonstrate that the natural motion in continuous space-time or
CSTM is not CCM, but one kind of essentially discontinuous motion,
we call it quantum discontinuous motion or quantum motion, since
we show that the wave function in quantum mechanics is the very
mathematical complex describing it, and Schr\"{o}dinger equation
of the wave function is also its simplest nonrelativistic motion
equation; Furthermore, since the combination of quantum mechanics
and general relativity will result in the discreteness of
space-time, namely the real space-time will be essentially
discrete, we further study the motion in discrete space-time, or
DSTM, and demonstrate that it will naturally result in the
collapse process of the wave function, and finally bring about the
appearance of CCM in macroscopic world.

The plan of this paper is as follows: In Sect. 2 we first give a
general analysis about CSTM, the motion state of particle is
physically defined, its general form and description are also
given based on the mathematical analysis in the Appendix. In Sect.
3 we work out the simplest evolution law of CSTM, which turns out
to be Schr\"{o}dinger equation in quantum mechanics. In Sect. 4 we
give a strict physical definition of CSTM, and further discuss the
constant $\hbar$ involved in its law. In Sect. 5 we point out that
space-time is essentially discrete due to the ubiquitous existence
of gravity, and give a simple demonstration. In Sect. 6 we further
give a general analysis about DSTM, and the general form of motion
state in such space-time is given. In Sect. 7 the evolution law of
DSTM is worked out, and we demonstrate that it will naturally
result in the collapse process of the wave function. In Sect. 8 we
further show that CCM and its evolution law can be consistently
derived from the evolution law of DSTM. At last, conclusions are
given.

\section{General analysis about motion in continuous space-time (CSTM)}
In this section, we will give a deep logical and physical analysis
about CSTM.

\subsection{The motion state of particle}
First, we should define the motion state of particle, there are
two alternatives, one is the instant state of particle, the other
is the infinitesimal interval state of particle, it has been
generally accepted that the motion state of particle should be the
infinitesimal interval state of particle, not the instant state of
particle, while people usually omit their essential difference,
here we will present some of them.

(1).The instant state of particle contains only one point in
space, its potential in mathematics is zero, while the
infinitesimal interval state of particle contains infinite
innumerable points in space, its potential in mathematics is
$\zeta_1$.

(2).The instant state of particle contains no motion, but only the
existence of particle, while the infinitesimal interval state of
particle may contain abundant motion elements, since it contains
infinite innumerable points in space.

(3).The instant state of particle possesses no physical meaning,
since we can not access it through physical measurement, while the
infinitesimal interval state of particle possesses real physical
meaning, since we can measure it by means of the following
infinite process:$\Delta{t}\rightarrow dt$.

(4).We can only find and confirm the law for the infinitesimal
interval state of particle, while as to the instant state of
particle, even if its law exists, we can not find it, let along
confirm it.

In fact, in physics there exist only the description quantities
defined during infinitesimal time interval, this fact can be seen
from the familiar differential quantities such as dt and dx,
whereas the quantities defined at instants come only from
mathematics, people always mix up these two kinds of quantities,
this is a huge obstacle for the development of physics. Thus we
can only discuss the motion state and relevant quantities defined
during infinitesimal time interval, as well as their differential
laws, if we study the point set corresponding to real physical
motion.

For simplicity, in the following we say the motion state of
particle at one instant, but it still denotes the infinitesimal
interval state of particle, not the instant state of particle.

\subsection{The general form of the motion state of particle}
Secondly, we will give the general form of the motion state of
particle, according to the analysis about point set(see Appendix),
the natural assumption in logic is that the motion state of
particle in infinitesimal time interval is a general dense point
set in space, since we have no a priori reason to assume a special
form, its proper description is the measure density $\rho(x,t)$
and measure density fluid $j(x,t)$.

Certainly, at some instant t the motion state of particle may
assume some kind of special form, such as the continuous point set
described by dx or $\rho(x,t)=\delta(x-x(t))$, but whether or not
this kind of special form can exist for other instants should be
determined by the motion law, not our prejudices.

\section{The evolution of motion in continuous space-time (CSTM)}

In the following, we will give the main clues for finding the
possible evolution equations of CSTM, and show that
Schr\"{o}dinger equation in quantum mechanics is just its simplest
nonrelativistic evolution equations. Here we mainly analyze
one-dimension motion, but the results can be easily extended to
three-dimension situation.

\subsection{The first motion principle}
First, we should find the first motion principle similar to the
first Newton principle, this means that we need to find the
simplest solution of the motion equation, in which we can get the
invariant quantity during free motion, it is evident that the
simplest solution of the motion equation is:
\begin{equation}\frac{\partial{\rho}(x,t)}{\partial{t}}=0\end{equation}
\begin{equation}\frac{\partial{j}(x,t)}{\partial{x}}=0\end{equation}
\begin{equation}\frac{\partial{j}(x,t)}{\partial{t}}=0\end{equation}
\begin{equation}\frac{\partial{\rho}(x,t)}{\partial{x}}=0\end{equation}
using the relation $j(x,t)$=$\rho(x,t)\cdot{v}$ we can further get
the solution, namely $\rho(x,t)$=1,$j(x,t)$=$v$=$p$/m, where $m$
is the mass of the particle, $p$ is defined as the momentum of
particle. Now, we get the first motion principle, namely during
the free motion of particle, the momentum of the particle is
invariant, but it can be easily seen that, contrary to classical
continuous motion, for the free particle with one constant
momentum, its position will not be limited in the infinitesimal
space dx, but spread throughout the whole space with the same
position measure density.

Similar to the quantity position, the natural assumption in logic
is also that the momentum (motion) state of particle in
infinitesimal time interval is still a general dense point set in
momentum space, thus we can also define the momentum measure
density $f(p,t)$, and the momentum measure fluid density $J(p,t)$,
their meanings are similar to those of position.

\subsection{Two kinds of description bases}

Now, we have two description quantities, one is position, the
other is momentum, and position descriptions $\rho(x,t)$ and
$j(x,t)$ provide a complete local description of the motion state,
we may call it local description basis, similarly momentum
descriptions $f(p,t)$ and $J(p,t)$ provide a complete nonlocal
description of the motion state, since for the particle with any
constant momentum, its position will spread throughout the whole
space with the same position measure density, we may call it
nonlocal description basis.

Furthermore, at any instant the motion state of particle is
unique, thus there should exist a one-to-one relation between
these two kinds of description bases, namely there should exist a
one-to-one relation between position description $(\rho,j)$ and
momentum description $(f,J)$, and this relation is irrelevant to
the concrete motion state, in the following we will mainly discuss
how to find this one-to-one relation, and our analysis will also
show that this relation essentially determines the evolution of
motion.

\subsection{One-to-one relation}

First, it is evident that there exists no direct one-to-one
relation between the measure density functions $\rho(x,t)$ and
$f(p,t)$, since even for the above simplest situation, we have
$\rho(x,t)=1$ and $f(p,t)=\delta^2(p-p_0)$\footnote{ This result
can be directly obtained when considering the general
normalization relation
$\int_{\Omega}\rho(x,t)dx=\int_{\Omega}f(p,t)dp$.}, and there is
no one-to-one relation between them.

Then in order to obtain the one-to-one relation, we have to create
new properties on the basis of the above position description
$(\rho,j)$ and momentum description $(f,J)$, this needs a little
mathematical trick, here we only give the main clues and the
detailed mathematical demonstrations are omitted, first, we
disregard the time variable $t$ and let $t=0$, as to the above
free evolution state with one momentum, we have
$(\rho,j)=(1,p_{0}/m)$ and $(f,J)=(\delta^2(p-p_0),0)$, thus we
need to create a new position state function $\psi(x,0)$ using $1$
and $p_{0}/m$, a new momentum state function $\varphi(p,0)$ using
$\delta^2(p-p_0)$ and $0$, and find the one-to-one relation
between these two state functions, this means there exists a
one-to-one transformation between the state functions $\psi(x,0)$
and $\varphi(p,0)$, we generally write it as follows:
\begin{equation}\label{}
\psi(x,0)=\int_{-\infty}^{+\infty}\varphi(p,0)T(p,x)dp
\end{equation}
where $T(p,x)$ is the transformation function and generally
continuous and finite for finite $p$ and $x$, since the function
$\varphi(p,0)$ will contain some form of the basic element
$\delta^2(p-p_0)$, normally we may expand it as
$\varphi(p,0)=\sum_{i=1}^{\infty}a_{i}\delta^{i}(p-p_0)$, while
the function $\psi(x,0)$ will contain the momentum $p_0$, and be
generally continuous and finite for finite $x$, then it is evident
that the function $\varphi(p,0)$ can only contain the term
$\delta(p-p_0)$, because the other terms will result in
infiniteness.

On the other hand, since the result $\varphi(p,0)=\delta(p-p_0)$
implies that there exists the simple relation
f(p,0)=$\varphi(p,0)^{*}\varphi(p,0)$\footnote{Evidently, another
simple relation f(p,0)=$\varphi(p,0)^{2}$ permit no existence of
one-to-one relation}, and owing to the equality between the
position description and momentum description, we also have the
similar relation $\rho(x,0)$=$\psi(x,0)^{*}\psi(x,0)$, thus we may
let $\psi(x,0)=e^{iG(p_{0},x)}$ and have $T(p,x)=e^{iG(p,x)}$,
then considering the symmetry between the properties position and
momentum\footnote{This symmetry essentially stems from the
equivalence between these two kinds of descriptions, the direct
implication is for $\rho(x,0)=\delta^2(x-x_0)$ we also have
$f(p,0)=1$.}, we have the general extension
$G(p,x)=\sum_{i=1}^{\infty}b_{i}(px)^{i}$, furthermore, this kind
of symmetry also results in the symmetry between the
transformation $T(p,x)$ and its reverse transformation
$T^{-1}(p,x)$, where $T^{-1}(p,x)$ satisfies the relation
$\varphi(p,0)=\int_{-\infty}^{+\infty}\psi(x,0)T^{-1}(p,x)dp$,
thus we can only have the term $px$ in the function $G(p,x)$, and
the resulting symmetry relation between these two transformations
will be $T^{-1}(p,x)=T^{*}(p,x)=e^{-ipx}$, we let $b_{1}=1/\hbar$,
where $\hbar$ is a constant quantity(for simplicity we let
$\hbar=1$ in the following discussions), then we get the basic
one-to-one relation, it is
$\psi(x,0)=\int_{-\infty}^{+\infty}\varphi(p,0)e^{ipx}dp$, where
$\psi(x,0)=e^{-ip_{0}x}$ and $\varphi(p,0)=\delta(p-p_0)$, it
mainly results from the essential symmetry involved in CSTM
itself.

In order to further find how the time variable $t$ is included in
the functions $\psi(x,t)$ and $\varphi(p,t)$, we may consider the
superposition of two single momentum states, namely
$\psi(x,t)=\frac{1}{\sqrt{2}}[e^{ip_{1}x-ic_{1}(t)}+e^{ip_{2}x-ic_{2}(t)}]$,
then the position measure density is $\rho(x,t)=[1+\cos(\triangle
c(t)-\triangle px)]/2$, where $\triangle c(t)=c_{2}(t)-c_{1}(t)$
and $\triangle p=p_{2}-p_{1}$, now we let $\triangle p\rightarrow
0$, then we have $\rho(x,t)\rightarrow 1$ and $\triangle
c(t)\rightarrow 0$, especially using the measure conservation
relation we can get $dc(t)/dt=dp\cdot p/m$, namely
$dc(t)=d(p^{2}/m)\cdot t$ or $dc(t)=dE\cdot t$, where $E=p^{2}/m$,
is defined as the energy of the particle in the nonrelativistic
domain, thus as to any single momentum state we have the
time-included formula $\psi(x,t)=e^{ipx-iEt}$.

In fact, there may exist other complex forms for the state
functions $\psi(x,t)$ and $\varphi(p,t)$, for example, they are
not the above simple number functions but multidimensional vector
functions such as
$\psi(x,t)=(\psi_{1}(x,t),\psi_{2}(x,t),...,\psi_{n}(x,t))$ and
$\varphi(p,t)=(\varphi_{1}(p,t),\varphi_{2}(p,t),...,\varphi_{n}(p,t))$,
but the above one-to-one relation still exists for every component
function, and these vector functions still satisfy the above
modulo square relations, namely
$\rho(x,t)=\sum_{i=1}^{n}\psi_{i}(x,t)^{*}\psi_{i}(x,t)$ and
$f(p,t)=\sum_{i=1}^{n}\varphi_{i}(p,t)^{*}\varphi_{i}(p,t)$, these
complex forms will correspond to the particles with more complex
structure, say, involving more inner properties of the particle
such as charge and spin etc.

At last, since the one-to-one relation between the position
description and momentum description is irrelevant to the concrete
motion state, the above one-to-one relation for the free motion
state with one momentum should hold true for any motion state,
and the states satisfying the one-to-one relation will be the
possible motion states. Furthermore, it is evident that this
one-to-one relation will directly result in the famous Heisenberg
uncertainty relation $\triangle x\cdot\triangle p\geq\hbar/2$.


\subsection{The evolution law of motion in continuous space-time (CSTM)}

Now, we will work out the evolution law of CSTM.

First, as to the free motion state with one momentum, namely the
single momentum state $\psi(x,t)=e^{ipx-iEt}$, using the above
definition of energy $E=\frac{p^{2}}{m}$ and including the
constant quantity $\hbar$ we can easily find its nonrelativistic
evolution law, which is
\begin{equation}
i\hbar\frac{\partial{\psi}(x,t)}{\partial{t}}=
-\frac{\hbar^2}{2m}\cdot{\frac{\partial^{2}\psi(x,t)}{\partial{x^2}}}
\end{equation}
then owing to the linearity of this equation, this evolution
equation also applies to the linear superposition of the single
momentum states, namely all possible free notion states, or we can
say, it is the free evolution law of CSTM.

Secondly, we will consider the evolution law of CTSM under outside
potential, when the potential $U(x,t)$ is a constant $U$, the
evolution equation will be
\begin{equation}
i\hbar\frac{\partial{\psi}(x,t)}{\partial{t}}=
-\frac{\hbar^2}{2m}\cdot{\frac{\partial^{2}\psi(x,t)}{\partial{x^2}}}
+U\cdot{\psi(x,t)}
\end{equation}
then when the potential $U(x,t)$ is related to x and t, the above
form will still hole true,
namely
\begin{equation}
i\hbar\frac{\partial{\psi}(x,t)}{\partial{t}}=
-\frac{\hbar^2}{2m}\cdot{\frac{\partial^{2}\psi(x,t)}{\partial{x^2}}}
+U(x,t)\cdot{\psi(x,t)}
\end{equation}
for three-dimension situation the equation will be
\begin{equation}
i\hbar\frac{\partial{\psi}(\mathbf{x},t)}{\partial{t}}=
-\frac{\hbar^2}{2m}\cdot\nabla^{2}\psi(\mathbf{x},t)
+U(\mathbf{x},t)\cdot{\psi(\mathbf{x},t)}
\end{equation}
this is just the Schr\"{o}dinger equation in quantum
mechanics\footnote{In fact, Schr\"{o}dinger equation is also the
simplest evolution equation of dense point set, this fact
manifests the identity of physical reality and mathematical
reality to some extent.}.

At last, the above analysis also shows that the state function
$\psi(x,t)$ provides a complete description of CSTM, since CSTM is
completely described by the measure density $\rho(x,t)$ and
measure fluid density $j(x,t)$, and according to the above
evolution equation the state function $\psi(x,t)$ can be expressed
by these two functions, namely
$\psi(x,t)=\rho^{1/2}\cdot{e^{iS(x,t)/\hbar}}$, where
$S(x,t)=m\int_{-\infty}^{x}
j(x^{'},t)/\rho(x^{'},t)dx^{'}+C(t)$\footnote{When in
three-dimension space, the formula for S(x,y,z,t) will be
$S(x,y,z,t)=m\int_{-\infty}^{x}
j_{x}(x^{'},y,z,t)/\rho(x^{'},y,z,t)dx^{'}+C(t)=m\int_{-\infty}^{y}
j_{y}(x,y^{'},z,t)/\rho(x,y^{'},z,t)dy^{'}+C(t)=m\int_{-\infty}^{z}
j_{z}(x,y,z^{'},t)/\rho(x,y,z^{'},t)dz^{'}+C(t)$, since in general
there exists the relation $\nabla \times
\{\mathbf{j}(x,y,z,t)/\rho(x,y,z,t)\}=0$.}, and these two
functions can also be expressed by the state function, namely
$\rho(x,t)=|\psi(x,t)|^2$ and $j(x,t) = [\psi^*
\partial \psi/\partial t - \psi \partial
 \psi^*/\partial t ]/2i$, then there exists a one-to-one relation
between $(\rho(x,t),j(x,t))$ and $\psi(x,t)$ when omitting the
absolute phase, thus the state function $\psi(x,t)$ also provides
a complete description of CSTM.

On the other hand, we can see that the absolute phase of the wave
function $\psi(x,t)$, which may depend on time, is useless for
describing CSTM, since according to the above analysis it
disappears in the measure density $\rho(x,t)$ and measure fluid
density $j(x,t)$, which completely describe CSTM, thus it is
natural that the absolute phase of the wave function possesses no
physical meaning.

\section{Further discussions about motion in continuous space-time (CSTM)}
\subsection{The definition of motion in continuous space-time (CSTM)}

Now we can give the physical definition of CSTM in three-dimension
space, the definition for other abstract spaces or many-particle
situation can be easily extended.

(1). The motion of particle in space is described by dense point
set in four-dimension space and time.

(2). The motion state of particle in space is described by the
position measure density $\rho(\mathbf{x},t)$ and position measure
fluid density $\mathbf{j}(\mathbf{x},t)$ of the corresponding
dense point set.

(3). The evolution of motion corresponds to the evolution of the
dense point set, and the simplest evolution equation is
Schr\"{o}dinger equation in quantum mechanics.

Compared with classical continuous motion, we may call CSTM
quantum discontinuous motion, or quantum motion, the commonness of
these two kinds of motion is that they are both the motion of
particle, namely the moving object exists only in one position in
space at one instant, their difference lies in the moving
behavior, namely the behavior of the particle during infinitesimal
time interval [t,t+dt], for classical motion, the particle is
limited in a certain local space interval [V,V+dV], while for
quantum motion, the particle moves throughout the whole space with
a certain position measure density $\rho(\mathbf{x},t)$.

In fact, all physical states of CSTM are defined during
infinitesimal time interval in the meaning of measure, not at one
instant, for example, the single momentum state
$\psi_{p}(x,t)=e^{ipx-iEt}$, especially even the single position
state $\psi(x,t)=\delta(x-x_0)$ is still defined during
infinitesimal time interval.

\subsection{Some discussions about the constant $\hbar$}

First, from the above analysis about CSTM, we can understand why
the constant $\hbar$ with dimension $J\cdot s$ should appear in
the evolution equation of CSTM, or Schr\"{o}dinger equation, the
existence of $\hbar$ essentially results from the equivalence
between the nonlocal momentum description and local position
description of CSTM, it is this equivalence that results in the
one-to-one relation between these two kinds of descriptions, which
requires the existence of a certain constant $\hbar$ with
dimension $J\cdot s$ to cancel out the dimension of the physical
quantities $px$ and $Et$ in the relation, at the same time, the
existence of $\hbar$ also indicates some kind of balance between
the dispersion of the position distribution and that of momentum
distribution limited by the one-to-one relation ( there is no such
limitation for CCM and $\hbar=0$ ), or we can say, the existence
of $\hbar$ essentially indicates some kind of balance between the
nonlocality and locality of motion in continuous space-time.

Secondly, even though the appearance of $\hbar$ in the evolution
equation of CSTM is inevitable, its value can not be determined by
CSTM itself, we only know that $\hbar$ possesses a finite nonzero
value, certainly, just like the other physical constants such as c
and G, its value can be determined by the experience, but there
may exist some deeper reasons for the special value of $\hbar$ in
our universe, although motion can not determine this value alone,
the solution may have to resort to other subtle realities in this
world, for example, gravity (G), space-time(c), or even the
existence of our mankind.

\section{The real space-time is discrete}

The combination of quantum mechanics and general relativity
strongly implied space-time is essentially discrete, and the
minimum space-time unit will be Planck size $T_{p}$ and $L_{p}$,
thus owing to the ubiquitous existence of gravity, the real
space-time will be essentially discrete with the minimum size
$T_{p}$ and $L_{p}$.

Here we will give a simple operational demonstration about the
discreteness of space-time, consider a measurement of the length
between points A and B, at point A place a clock with mass $m$ and
size $a$ to register time, at point B place a reflection mirror,
when t=0 a photon signal is sent from A to B, at point B it is
reflected by the mirror and returns to point A, then the clock
registers the return time, for classical situation the measured
length will be $L=\frac{1}{2}ct$, but when considering quantum
mechanics and general relativity, the existence of the clock
introduces two kinds of uncertainties to the measured length, the
uncertainty resulting from quantum mechanics is: $\delta
L_{QM}\geq (\frac{\hbar L}{mc})^{1/2}$, the uncertainty resulting
from general relativity is: $\delta L_{GR} \cong
\frac{1}{2}\frac{2Gm}{c^{2}}ln\frac{a+L}{a} \geq \frac{Gm
}{c^{2}}$, then the total uncertainties is:$\delta L=\delta
L_{QM}+\delta L_{GR} \geq (L\cdot{L_{p}}^{2})^{1/3}$, where
$L_{p}=(\frac{G\hbar}{c^{3}})^{1/2}$, is Planck length, thus we
conclude that the minimum measurable length is Planck length
$L_{p}$, in a similar way, we can also work out the minimum
measurable time, it is just Planck time $T_{p}$.

\section{General analysis about motion in discrete space-time (DSTM)}

In the discrete space-time, there exist absolute minimum sizes
$T_{p}$ and $L_{p}$, namely the minimum distinguishable size of
time and position of the particle is respectively $T_{p}$ and
$L_{p}$, thus in physics the existence of the particle is no
longer in one position at one instant as in the continuous
space-time, but limited in a space interval $L_{p}$ during a
finite time interval $T_{p}$, it can be seen that this state
corresponds to the instant state of particle in continuous
space-time, we define it as the instant state of particle in
discrete space-time, this state evidently contains no motion, but
only the existence of particle.

Furthermore, during the finite time interval $T_{p}$ the particle
can only be limited in a space interval $L_{p}$, since if it can
move throughout at least two different local regions with
separation size larger than $L_{p}$ during the time interval
$T_{p}$, then there essentially exists a smaller distinguishable
finite time interval than $T_{p}$, which evidently contradicts the
fact that $T_{p}$ is the minimum time unit, thus the discreteness
of space-time essentially results in the existence of local
position state of the particle, in which the particle stays in a
local region with size $L_{p}$ for a time interval $T_{p}$, we may
call such general local position state Planck cell state.

Similar to the analysis of the motion state in continuous
space-time, in discrete space-time, the natural assumption in
logic is that the motion state of particle in finite time
interval, which is much longer than $T_{p}$ but still small
enough, is a general discrete point set or cell set in space,
since we have no a priori reason to assume a special form, its
proper description is still the measure density $\rho(x,t)$ and
measure density fluid $j(x,t)$, but in the
meaning of time average.

Certainly, we can also get the motion state of particle in
discrete space-time from that in continuous space-time, since in
continuous space-time the particle, which instant state is the
particle being in one position at one instant, moves throughout
the whole space during infinitesimal time interval, while in
discrete space-time the instant state of particle turns to be the
particle being in a space interval $L_{p}$ during a finite time
interval $T_{p}$, the motion state of particle in discrete
space-time will naturally be that during a finite time interval
much larger than $T_{p}$ the particle moves throughout the whole
space with the position measure density $\rho(x,t)$ in the meaning
of time average.

Now the visual physical picture of DSTM will be that during a
finite time interval $T_{p}$ the particle will stay in a local
region with size $L_{p}$, then it will still stay there or "jump"
to another local region, which may be very far from the original
region, while during a time interval much larger than $T_{p}$ the
particle will move throughout the whole space with a certain
average position measure density $\rho(x,t)$.

As we can see, on the one hand, the particle undergoing DSTM stays
in a local region during infinitesimal time interval, this may
generate the display of CCM, on the other hand, during a finite
time interval much larger than $T_{p}$ the particle will
continually jumps from one local region to another local region,
and move throughout the whole space with a certain position
measure density $\rho(x,t)$, this may generate the display of
CSTM, then DSTM is evidently some kind of unification of CCM and
CSTM, in fact, owing to the ubiquitous existence of quantum and
gravity, space-time is essentially discrete, and DSTM will be the
only real motion in Nature, while CSTM and CCM are the apparent
motion modes, they are only two ideal approximators of DSTM in
microscopic and macroscopic world, thus DSTM is just the lost
reality unifying quantum motion (CSTM) and classical motion (CCM),
and it will undoubtedly provide an uniform realistic picture for
microscopic world and macroscopic world, the following analysis
will confirm this conclusion more convincingly.

\section{The evolution of motion in discrete space-time (DSTM)}

\subsection{A general discussion}
Since CSTM is some kind of time average of DSTM, the evolution of
DSTM will follow the evolution law of CSTM in the meaning of time
average, on the other hand, the particle undergoing DSTM does stay
in a local region for a finite nonzero time interval, and jump
from this local region to another local region stochastically,
thus the position measure density $\rho(x,t)$ of the particle will
be essentially changed in a stochastic way due to the finite
nonzero stay time in different stochastic region\footnote{For
CSTM, the stay time of the particle in any position is zero, so
its position measure density $\rho(x,t)$ is not influenced by the
stochastic jump.}, and the corresponding wave function will be
also stochastically changed, then this kind of stochastic jump
inevitably introduce the stochastic element to the evolution, so
the evolution law of DSTM will be the combination of the
deterministic linear evolution and stochastic nonlinear evolution,
in the following we will work out this law.

\subsection{Two rules}
At first, since CSTM is some kind of average of DSTM during a
finite time interval much larger than $T_{p}$, thus the position
measure density $\rho(x,t)$ of the particle undergoing CSTM will
be also the average of the position distribution of the particle
undergoing DSTM during this time interval, then it is natural that
the position of the particle undergoing DSTM will satisfy the
position measure density $\rho(x,t)$, namely for DSTM the
stochastic stay position of the particle satisfies the
distribution
\begin{equation}\label{}
P(x,t)=|\psi(x,t)|^{2}
\end{equation}
this is the first useful rule for finding the evolution law of
DSTM.

Secondly, according to the definition of the position measure
density $\rho(x,t)$, the finite nonzero stay time of the particle
in a local region evidently implies that the position measure
density $\rho(x,t)$ in that region will be increased after this
finite nonzero stay time interval, and the increase will be larger
when the stay time is longer. We consider the general situation
that the particle undergoing DSTM stays in a local region $L_{p}$
for a time interval $T$, in the first rank approximation the
increase of the position measure density $\rho(x,t)$ in this
region can be written as follows after normalization:
\begin{equation}\label{}
\rho(x,t+T)=\frac{1}{A(T)}(\rho(x,t)+T/T_m)
\end{equation}
where $A(T)$ is the normalization factor, $T_m$ is a certain time
size to be determined, which may be relevant to the concrete
motion state of the particle, this will be the second useful rule.

We first work out the normalization factor $A(t)$, considering the
following two conditions:(1)when $T=0$,$\rho(x,t+T)=\rho(x,t)$,
and $A(0)=1$;(2)when $T\rightarrow \infty$,$\rho(x,t+T)\rightarrow
1$, and $A(\infty)\rightarrow T/T_m$, we can get $A(T)=1+T/T_m$,
then the above formula will be:
\begin{equation}\label{}
\rho(x,t+T)=\frac{\rho(x,t)+T/T_m}{1+T/T_m}
\end{equation}
or it can be written as follows:
\begin{equation}\label{}
\Delta\rho(x,t)=\frac{T}{T_m+T}(1-\rho)
\end{equation}
in general, when $T>T_p$, namely when the particle undergoing DSTM
stays in a local region $L_{p}$ for a time interval longer than
$T_p$, we can divide the whole time interval $T$ into many Planck
cell $T_p$, and the above formula is still valid for every cell,
thus in the following discussions we let $T=T_p$ for simplicity.

In order to further find the formula of $T_m$, we need to study
the limitation on the jump of the particle undergoing DSTM, since
the item $T_m$ just denotes this kind of limitation, this can be
seen from the following two extreme situations:(1)when
$T_m\rightarrow \infty$, we have $\Delta \rho\rightarrow 0$, this
denotes that the position measure density will be not influenced
by the jump, and the particle can jump freely;(2)$T_m\rightarrow
0$, we have $\rho\rightarrow 1$, this denotes that the position
measure density will turn to be one in the region where the
particle stays, and the position measure density in other regions
will turn to be zero, so the particle can not jump at all. In
fact, from the physical analysis about the jump we can see that
the limitation results from the principle of energy conservation,
according to which during a finite nonzero time interval $\Delta
t$ the possible change of energy $\Delta E_j$ resulting from jump
will be limited by the uncertainty relation $\Delta E_j \approx
\hbar / \Delta t$, now we consider two situations, first, if the
total difference of energy $\Delta E$ between the original stay
region and other regions satisfies the condition $\Delta E \gg
\Delta E_{j}$, then the particle can hardly jump from its original
region to other regions,
namely after the stay time $\Delta t$ the position measure density
$\rho(x,t)$ in the original region will be greatly increased,
especially when $\Delta E \rightarrow \infty$, we have
$\rho(x,t)\rightarrow 1$, and $T_m\rightarrow 0$\footnote{In fact,
in this situation the wave function has collapsed into this local
region in order to satisfy the requirement of energy conservation,
and this also indicates that in order to satisfy the principle of
energy conservation DSTM will naturally result in the collapse of
the wave function.}; Secondly, if the total difference of energy
$\Delta E$ between the original stay region and other regions
satisfies the condition $\Delta E \ll \Delta E_{j}$, then the
particle can jump more easily from its original region to other
regions, namely after the stay time $\Delta t$ the position
measure density $\rho(x,t)$ will be only changed slightly,
especially when $\Delta E \rightarrow 0$, we have
$\Delta\rho(x,t)\rightarrow 0$, and $T_m\rightarrow \infty$. Then
we can see that $T_m$ is inversely proportional to $\Delta E$,
considering the dimension requirement their relation will be
$T_m=\hbar/k\Delta E$, where $k$ is a dimensionless constant.

Now, the change of the position measure density after stay time
$T_p$ can be formulated in a more complete way:
\begin{equation}\label{}
\rho(x,t+T)=\frac{\rho(x,t)+k\Delta E/E_{p}}{1+k\Delta E/E_{p}}
\end{equation}
or it can be written as follows:
\begin{equation}\label{}
\Delta\rho(x,t)=\frac{\Delta E}{kE_{p}+\Delta E}(1-\rho)
\end{equation}
where $E_{p}=\hbar/T_p$ is Planck energy, thus we get the second
useful rule for finding the evolution law of DSTM.

\subsection{The evolution law of motion in discrete space-time (DSTM)}

Now, according to the above two rules, we can give the evolution
equation of DSTM.

For simplicity but lose no generality, we consider a one-dimension
initial wave function $\psi(x,0)$, according to the above
analysis, the concrete evolution equation of DSTM will be
essentially one kind of revised stochastic evolution equation
based on Schr\"{o}dinger equation, here we assume the form of
stochastic differential equation ( SDE ), it can be written as
follows:
\begin{equation}\label{}
d\psi(x,t)=\frac{1}{i\hbar}H_{Q}\psi(x,t)dt+\frac{1}{2}[\frac{\delta_{xx_{N}}}{\rho(x,t)}-1]
\frac{\Delta E(x_{N},\overline{x_{N}})}{kE_{p}+\Delta
E(x_{N},\overline{x_{N}})}\psi(x,t)dt
\end{equation}
where the first term in right side represents the evolution
element resulting from CTSM, the average behavior of DSTM, $H_Q$
is the corresponding Hamiltonian, the second term in right side
represents the evolution element resulting from the stochastic
jump resulting from DSTM itself, $\delta_{xx_{N}}$ is the discrete
$\delta$-function, $k$ is a dimensionless constant,
$\rho(x,t)=|\psi(x,t)|^{2}$, is the position measure density,
$\Delta E(x_{N},\overline{x_{N}})$ is the total difference of
energy of the particle between the cell containing $x_{N}$ and all
other cells $\overline{x_{N}}$, $x_{N}$ is a stochastic position
variable, whose distribution is $P(x_{N},t)=\rho(x_{N},t)$.

In physics, this stochastic differential equation is essentially a
discrete evolution equation, all the quantities are defined
relative to the Planck cells $T_{p}$ and $L_{p}$, and the equation
should be also solved in a discrete way.

\subsection{Some further discussions}
Now we will give some physical analyses about the above evolution
equation of DSTM, first, the linear item in the equation will
result in the spreading process of the wave function as for the
evolution of CSTM, while the nonlinear stochastic item in the
equation will result in the localizing process of the particle or
collapse process of the wave function, this can also be seen
qualitatively, since according to the nonlinear stochastic item,
in the region where the position measure density is larger the
stay time of the particle will be longer, moreover, the longer
stay time of the particle in one region will further increase the
position measure density in that region much more, thus this
process is evidently one kind of positive feedback process, the
particle will finally stay in a local region, and the wave
function of particle will also collapse to that region, so the
evolution of DSTM will be some kind of combination of the
spreading process and localizing process.

Secondly, the strength of the spreading process and localizing
process is mainly determined by the energy difference between
different branches of the wave function, if the energy difference
is so small, then the evolution of DSTM will be mainly dominated
by the spreading process, or we can say, the display of DSTM will
be more like that of quantum motion (CSTM), this is just what
happens in microscopic world; while if the energy difference is so
large, then the evolution of DSTM will be mainly dominated by the
localizing process, or we can say, the display of DSTM will be
more like that of classical motion (CCM), this is just what
happens in macroscopic world, and the boundary of these two worlds
can also be estimated, the following example indicates that the
energy difference in the boundary may assume $\Delta
E\cong\sqrt{\hbar E_{p}} \approx 7Mev$, the corresponding collapse
time will be in the level of seconds.

Thirdly, if the particle finally stay in a local region during the
evolution of DSTM, the localizing probability of the particle, or
the collapse probability the wave function in one local region is
just the initial position measure density of the particle in that
region, namely the probability satisfies the Born rule in quantum
mechanics, since the stochastic evolution of DSTM satisfies the
Martingale condition, this can be seen from the following fact,
namely during every jump the position measure density $\rho$
satisfies the equation $P(\rho)=\rho P(\rho+\frac{\Delta E}{k
E_{p}+\Delta E}(1-\rho))+(1-\rho)P(\rho-\frac{\Delta E}{k
E_{p}+\Delta E}\rho)$\cite{Pea}, where $P(\rho)$ is the
probability of $\rho$ turning into one in one local region, namely
the probability of the particle localizing in a local region,
moreover, the solution of this equation is $P(\rho)=\rho$, this
just means that the localizing probability of the particle in one
region is just the initial position measure density of the
particle in that region.

Fourthly, the collapse process resulting from the evolution of
DSTM has no tails, since the evolution is essentially discrete,
the wave function is just the description of the motion of
particle, and its existence is only in the meaning of time
average, while the particle, the real object, always exists in one
local position, thus in the last stage of the collapse process,
when the particle stays in one of the branches long enough it will
de facto collapse into that branch owing to the limitation of
energy conservation, and the wave function, the apparent "object",
will also completely disappears in other branches\footnote{If the
wave function is taken as some kind of essential existence, and
its evolution is essentially continuous, then the tails problem
will be inevitable.}.

At last, the existence of DSTM will help to tackle the well-known
time problem involved in formulating a complete theory of quantum
gravity\cite{Pen}, since as to DSTM, the local position state of
particle will be the only proper state, and the only real physical
existence, during a finite time interval $T_{p}$ the particle can
only be limited in a local space interval $L_{p}$, namely there
does not exist any essential superposition of different positions
at all, the superposition of the wave function is only in the
meaning of time average, thus the essential inconsistency of the
superposition of different space-time in the theory of quantum
gravity, which results from the existence of the essential
superposition of the wave function, will naturally disappear, and
the real physical picture based on DSTM will be that at any
instant ( during a finite time interval $T_{p}$ ) the structure of
space-time determined by the existence of the particle ( in a
local space interval $L_{p}$ ) is definite or "classical", while
during a finite time interval much larger than $T_{p}$ but still
small enough it will be stochastically disturbed by the stochastic
jump of the particle undergoing DSTM, this kind of stochastic
disturbance will be the real quantum nature of the space-time and
matter.

\subsection{One simple example}
In this section, as one example we will analyze the DSTM evolution
of a simple two-state system, and quantificationally show that the
evolution of DSTM will indeed result in the collapse process of
the wave function.

We suppose the initial wave function of the particle is
$\psi(x,0)=\alpha(0)^{1/2} \psi_{1}(x)+\beta(0)^{1/2}
\psi_{2}(x)$, which is a superposition of two static states with
different energy levels $E_{1}$ and $E_{2}$, these two static
states are located in separate regions $R_{1}$ and $R_{2}$ with
the same size.

Since the energy of the particle inside the region of each static
state is the same, we can consider the spreading space of both
static states as a whole local region, and only study the
stochastic jump between these two regions resulting from the
evolution of DSTM, namely we directly consider the difference of
the energy $\Delta E=E_{2}-E_{1}$ between these two states,
through some mathematical calculations we can work out the density
matrix of the two-state system, it is:
\begin{equation}\label{}
\rho_{11}(t)=\alpha(0)
\end{equation}
\begin{equation}\label{}
\rho_{22}(t)=\beta(0)
\end{equation}
\begin{equation}\label{}
\rho_{12}(t)=[1-(\frac{\Delta E }{k E_{p} +\Delta E
})^{2}]^{t/2T_{p}}\sqrt{\alpha(0)\beta(0)} \approx(1-\frac{(\Delta
E)^{2}}{2k^{2}\hbar E_{p}}t)\sqrt{\alpha(0)\beta(0)}
\end{equation}
\begin{equation}\label{}
\rho_{21}(t)=[1-(\frac{\Delta E }{k E_{p} +\Delta
E})^{2}]^{t/2T_{p}}\sqrt{\alpha(0)\beta(0)}
\approx(1-\frac{(\Delta E)^{2}}{2k^{2}\hbar
E_{p}}t)\sqrt{\alpha(0)\beta(0)}
\end{equation}

It is evident that these results confirm the above qualitative
analysis definitely, namely, the evolution of DSTM indeed results
in the collapse of the wave function describing DSTM, and the
distribution of the collapse results satisfies the Born rule in
quantum mechanics, besides, we also get the concrete collapse time
for two-state system, it is $\tau_{c} \approx 2k^{2}\frac{\hbar
E_{p}}{(\Delta E)^{2}}$\footnote{This result has also been
obtained by Hughston\cite{Hughston} and Fivel\cite{Fivel} from
different point of views, and discussed by Adler et
al\cite{Adler1,Adler2}.}.

\section{The appearance of classical motion in macroscopic world}
The above analysis has indicated that, when the energy difference
between different branches of the wave function is large enough,
say for the macroscopic situation\footnote{The largeness of the
energy difference for macroscopic object results mainly from the
environmental influences such as thermal energy fluctuations.},
the linear spreading of the wave function will be greatly
suppressed, and the evolution of the wave function will be
dominated by the localizing process, in fact, the motion state of
the particle will be only local position state in appearance, and
the evolution of this state will be only still or continuously
move in space, this is just the display of CCM in macroscopic
world.

Furthermore, we will show that the evolution law of CCM can also
be derived, in fact, some people have strictly given the
demonstration based on revised quantum dynamics
\cite{Ghir86,Ghir90}, here we simply use the Enrenfest theorem,
namely $\frac{d<x>}{dt}=<p>$ and $\frac{d<p>}{dt}=<-\frac{\partial
U}{\partial x}>$, as we have demonstrated, for macroscopic object
its wave function will no longer spread, thus the average items in
the theorem will represent the effective description quantities
for the classical motion of the macroscopic object, and the
classical motion law is also naturally derived in such a way, the
result is $\frac{dx}{dt}=p$, the definition of the momentum, and
$\frac{dp}{dt}=-\frac{\partial U}{\partial x}$, the motion
equation.

\section{Conclusions}
In this paper, we strictly demonstrate the logical inevitability
of the existent form and evolution law of CSTM, the existence of
discrete space-time in Nature and resulting real existence of DSTM
and its evolution law, this not only explains the appearance of
classical motion in macroscopic world, as well as quantum motion
in microscopic world consistently and objectively, but also
presents a clear logical connection between quantum motion and
classical motion, and unveils the unified realistic picture of
microscopic and macroscopic world.

\vskip .5cm \noindent Acknowledgments \vskip .5cm Thanks for
helpful discussions with X.Y.Huang ( Peking University ),
A.Jadczyk ( University of Wroclaw ), P.Pearle ( Hamilton College
), F.Selleri ( University di Bari ), Y.Shi ( University of
Cambridge ), A.Shimony, A.Suarez ( Center for Quantum Philosophy
), L.A.Wu ( Institute Of Physics, Academia Sinica ), Dr S.X.Yu (
Institute Of Theoretical Physics, Academia Sinica ), H.D.Zeh.

\vskip 1cm \centerline{Appendix:Mathematical Analysis About Motion
In Continuous Space-time} \vskip .5cm

First, we will give three general presuppositions about the
relation between physical motion and mathematical point set, they
are basic conceptions and correspondence rules before we discuss
the physical motion of particles in continuous space-time.

(1). Time and space in which the particle moves are both
continuous point set.

(2). The moving particle is represented by one point in time and
space.

(3). The motion of particle is represented by the point set in
time and space.

The first presupposition defines the continuity of space-time, the
second one defines the existent form of particle in time and
space, the last one relates the physical motion of particle with
the mathematical point set.

For simplicity but lose no generality, in the following we will
mainly analyze the point set in two-dimension space-time, which
corresponds to one-dimension motion in continuous space-time.

\subsection{Point set and its law---a general discussion}
As we know, the point set theory has been deeply studied since the
beginning of this century, nowadays we can grasp it more easily,
according to this theory, we know that the general point set is
dense point set, whose basic property is the measure of the point
set, while the continuous point set is one kind of special dense
point set, its basic property is the length of the point set.

Naturally, as to the point set in two-dimension space-time, the
general situation is the dense point set in this two-dimension
space-time, while the continuous curve is one kind of extremely
special dense point set, surely it is a wonder that so many points
bind together to form one continuous curve by order, in fact, the
probability for its natural formation is zero.

Now, we will generally analyze the law of the point set, as we
know, the law about the points in point set, which can be called
point law, is the most familiar law, and it is widely taken as the
only rational law, for example, as to the continuous curve in
two-dimension space-time there may exist a certain expressible
analytical formula for the points in this special point
set\footnote{People cherish this kind of point laws owing to their
infrequent existence, but perhaps Nature detests and rejects them,
since the probability of creating them is zero.}, but as to the
general dense point set in two-dimension space-time the point law
possesses no mathematical meaning, since the dense point set is
discontinuous everywhere, even if the difference of time is very
small, or infinitesimal, the difference of space can be very
large, then infinitesimal error in time will result in finite
error in space, thus even if the point law exists we can not
formulate it in mathematics, and owing to finite error in time
determination and calculation, we can not prove it either, let
alone predict the evolution of the point set using it, in one
word, there does not exist point law for dense point set in
mathematics.

\subsection{Deep analysis about dense point set}
Now, we will further study the differential description of point
set in detail.

First, in order to find the differential description of the
special dense point set---continuous curve, we may measure the
rise or fall size of the space $\Delta{x}$ corresponding to any
finite time interval $\Delta{t}$ near each instant $t_j$, then at
any instant $t_j$ we can get the approximate information about the
continuous curve through the quantities $\Delta{t}$ and
$\Delta{x}$ at that instant, and when the time interval
$\Delta{t}$ turns smaller, we will get more accurate information
about the curve. In theory, we can get the complete information
through this infinite process, that is to say, we can build up the
basic description quantities for the special dense point
set---continuous curve, which are the differential quantities dt
and dx, then given the initial condition the relation between dt
and dx at all instants will completely describe the continuous
curve.

Then, we will analyze the differential description of the general
dense point set, as to this kind of point set, we still need to
study the concrete situation of the point set corresponding to
finite time interval near every instant. Now, when time is during
the interval $\Delta{t}$ near instant $t_j$, the points in space
are no longer limited in the local space interval $\Delta{x}$,
they distribute throughout the whole space instead, so we should
study this new point set, which is also dense point set, for
simplicity but lose no generality, we consider finite space such
as x$\in$[0,1], first, we may divide the whole space in small
equal interval $\Delta{x}$, the dividing points are denoted as
$x_i$, then we can define and calculate the measure of the local
dense point set in the space interval $\Delta{x}$ near each $x_i$,
which can be written as $M_{\Delta{x},\Delta{t}}$($x_i$,$t_j$),
since the measure sum of all local dense point sets in time
interval $\Delta{t}$ just equals to the length of the continuous
time interval $\Delta{t}$, we have:
\begin{equation}\sum_{i}M_{\Delta{x},\Delta{t}}$($x_i$,$t_j$)=
$\Delta{t}\end{equation}

On the other hand, since the measure
$M_{\Delta{x},\Delta{t}}$($x_i$,$t_j$) will also turn to be zero
when the intervals $\Delta{x}$ and $\Delta{t}$ turn to be zero, it
is not an useful quantity, and we have to create a new quantity on
the basis of this measure. Through further analysis, we find that
a new quantity
$\rho_{\Delta{x},\Delta{t}}(x_i,t_j)=M_{\Delta{x},\Delta{t}}(x_i,t_j)/(\Delta{x}\cdot\Delta{t})$,
which can be called average measure density, will be an useful
one, it generally does not turn to be zero when $\Delta{x}$ and
$\Delta{t}$ turn to be zero, especially if the limit
$lim_{\Delta{x}\rightarrow0,\Delta{t}\rightarrow0}\rho_{\Delta{x},\Delta{t}}(x_i,t_j)$
exists, it will no longer relate to the observation sizes
$\Delta{x}$ and $\Delta{t}$, so it can accurately describe the
whole dense point set, as well as all local dense point sets near
every instant, now we let:
\begin{equation}
\rho(x,t)=lim_{\Delta{x}\rightarrow0,\Delta{t}\rightarrow0}\rho_{\Delta
{x},\Delta{t}}(x,t)\end{equation} then we can get:
\begin{equation}
\int_{\Omega}\rho(x,t)dx=1
\end{equation}
this is just the normalization formula, where $\rho$(x,t) is
called position measure density, $\Omega$ denotes the whole
integral space.

Now, we will analyze the new quantity $\rho$(x,t) in detail,
first, the position measure density $\rho$(x,t) is not a point
quantity, it is defined during infinitesimal interval, this fact
is very important, since it means that if the measure density
$\rho(x,t)$ exists, then it will be continuous relative to both t
and x, 
that is to say, contrary to the position function x(t), there does
not exist the discontinuous situation for the measure density
function $\rho$(x,t), furthermore, this fact also results in that
the continuous function $\rho$(x,t) is the last useful quantity
for describing the dense point set; Secondly, the essential
meaning of the position measure density $\rho$(x,t) lies in that
it represents the dense degree of the points in the point set in
two-dimension space and time, and the points are denser where the
position measure density $\rho$(x,t) is larger.

\subsection{The evolution of dense point set}
Now, we will further discuss the evolution law for dense point
set.

Just like the continuous position function $x(t)$, although the
continuous position measure density function $\rho$(x,t)
completely describes the dense point set, it is one kind of static
description about the point set, and it can not be used for
prediction itself, so in order to predict the evolution of the
dense point set we must create some kind of quantity describing
its change, enlightened by the theory of fluid mechanics we can
define the fluid density for the position measure density
$\rho$(x,t) as follows:
\begin{equation}
\frac{\partial{\rho}(x,t)}{\partial{t}}+\frac{\partial{j}(x,t)}{\partial{x}}
=0
\end{equation}
we call this new quantity $j(x,t)$ position measure fluid density,
this equation measure conservation equation, it is evident that
this quantity just describes the change of the measure density of
dense point set, thus the general evolution equations of dense
point set can be written as in the following:
\begin{equation}
\frac{\partial{\rho}(x,t)}{\partial{t}}+\frac{\partial{j}(x,t)}{\partial{x}}=0
\end{equation}
\begin{equation}
\frac{\partial{j(x,t)}}{\partial{t}}+...=0
\end{equation}

\end{document}